\documentclass[conference]{IEEEtran}
 \IEEEoverridecommandlockouts
 
 \usepackage{mathrsfs}%flower
 \usepackage{amsmath}
 \usepackage{ntheorem}
 
 \usepackage{makeidx}  % allows for indexgeneration
 \usepackage{algorithm}
 \usepackage{algorithmic}
 \usepackage{graphicx}
 \usepackage{subfigure}
 \usepackage{epstopdf}
 \usepackage{bm}
 \usepackage{cite}
 \usepackage{stfloats}

 \usepackage{amssymb}
 \usepackage{graphicx}
 \usepackage{url}
 
 \usepackage{tabularx,booktabs}
 \newcolumntype{C}{>{\centering\arraybackslash}X} % centered version of "X" type
 \usepackage{lipsum}
 
 \usepackage{makecell} % Xhline Xcline
 
 \usepackage{graphicx}
 \usepackage{color}
\renewcommand{\algorithmicrequire}{\textbf{Input:}} 
\renewcommand{\algorithmicensure}{\textbf{Output:}}

 \usepackage{diagbox}
 
 \def\BibTeX{{\rm B\kern-.05em{\sc i\kern-.025em b}\kern-.08em
 		T\kern-.1667em\lower.7ex\hbox{E}\kern-.125emX}}

\begin{document}
\theoremseparator{.}

\newtheorem{lemma}{\bf Lemma }
\newtheorem{theorem}{\bf Theorem}
\newtheorem{corollary}{Corollary}
\renewcommand{\algorithmicrequire}{\textbf{Input:}} 
\renewcommand{\algorithmicensure}{\textbf{Output:}}
\newtheorem{remark}{Remark}
\title {Device Scheduling for Over-the-Air Federated Learning with Differential Privacy}

	\author{Na Yan\textsuperscript{1}, Kezhi Wang\textsuperscript{2}, Cunhua Pan\textsuperscript{3}, and Kok Keong Chai\textsuperscript{1}\\
	\textsuperscript{1}School of Electronic Engineering and Computer Science, Queen Mary University of London, U.K.  \\
	\textsuperscript{2}Department of Computer Science, Brunel University London, U.K. \\
	\textsuperscript{3}National Mobile Communications Research Laboratory, Southeast University, China.	\\
	Email: \{n.yan, michael.chai\}@qmul.ac.uk, kezhi.wang@brunel.ac.uk, and cpan@seu.edu.cn
	\thanks{This work of Na Yan was supported by China Scholarship Council.}
}
\maketitle

\begin{abstract}
	In this paper, we propose a device scheduling scheme for differentially private over-the-air federated learning (DP-OTA-FL) systems, referred to as S-DPOTAFL, where the privacy of the participants is guaranteed by channel noise. 
	In S-DPOTAFL, the gradients are aligned by the alignment coefficient and aggregated via over-the-air computation (AirComp). The scheme schedules the devices with better channel conditions in the training to avoid the problem that the alignment coefficient is limited by the device with the worst channel condition in the system. We conduct the privacy and convergence analysis to theoretically demonstrate the impact of device scheduling on privacy protection and learning performance. To improve the learning accuracy, we formulate an optimization problem with the goal to minimize the training loss subjecting to privacy and transmit power constraints. Furthermore, we present the condition that the S-DPOTAFL performs better than the DP-OTA-FL without considering device scheduling (NoS-DPOTAFL). The effectiveness of the S-DPOTAFL is validated through simulations.
\end{abstract}


\IEEEpeerreviewmaketitle
\section{Introduction}
Federated learning (FL) \cite{mcmahan2017communication} is a distributed machine learning (ML) framework whose goal is to train high-quality ML models without compromising users' data privacy. In FL, edge devices train a shared ML model collaboratively using their local data, with the help of a central controller, such as a base station (BS), which updates the global model and coordinates the training process. By training models locally, FL not only makes full use of the edge devices' computing power but also reduces the power consumption, latency, and privacy exposure due to the transmission of raw data. However, despite these promising benefits, FL still has issues of privacy leakage \cite{melis2019exploiting,zhu2019deep} and large upload latency\cite{kairouz2021advances,mcmahan2017communication}.
%However, despite these promising benefits, FL involves the following two challenges: First, although FL avoids the direct exposure of the user's sensitive information, it still faces the risk of privacy leakage if attacking the exchanged messages \cite{melis2019exploiting,zhu2019deep}. Second, the massive connectivities for uploading gradients via resource-limited wireless multiple access channel (MAC) result in considerable upload latency \cite{kairouz2021advances,mcmahan2017communication}.
% This happens because the model parameters uploaded by clients may still be informative [3]?[5], and the clients? private data may be inferred by malicious central servers from the aggregated global model parameters [5].

Over-the-air FL (OTA-FL) was introduced to reduce the communication latency where the gradients are transmitted via analog signals, which avoids the latency due to the quantization and encoding/decoding \cite{amiri2020machine}. In addition, it is highly bandwidth efficient compared with the traditional communication-and-computation separation method since the bandwidth allocation is independent of the massive number of devices \cite{nazer2007computation,goldenbaum2013harnessing}. By applying differential privacy (DP) \cite{dwork2014algorithmic} on OTA-FL, referred to as DP-OTA-FL, the two challenges aforementioned can be overcome simultaneously. 

%Over-the-air FL (OTA-FL) \cite{yang2020federated} with differential privacy (DP)  \cite{dwork2014algorithmic}, referred to as DP-OTA-FL, is a promising paradigm to overcome the two challenges jointly. On the one hand, DP prevents privacy leakage of FL by introducing random noise into the disclosed gradients to mask the contribution of any individual data point. With OTA-FL, the gradients are transmitted simultaneously via a shared wireless multiple access channel (MAC) in an analog way and aggregated ``over-the-air" thanks to the waveform-superposition property. First of all, OTA-FL can reduce upload latency by avoiding quantization and encoding/decoding \cite{amiri2020machine}. Furthermore, OTA-FL is highly bandwidth efficient compared with the traditional communication-and-computation separation method since the bandwidth allocation is independent of the massive number of devices . 

To mitigate the impact of fading wireless channels on transmitting gradients, the existing works on DP-OTA-FL considered aligned OTA-FL \cite{seif2020wireless,koda2020differentially,liu2020privacy}, where the gradients are aligned by adjusting the channel conditions of all devices equal to an alignment coefficient. In these works, artificial noise and channel noise were employed to enhance privacy. In \cite{seif2020wireless}, artificial Gaussian noise was added to each gradient before transmitting if channel noise cannot provide sufficient privacy protection. The amount of the artificial noise was calculated based on the alignment coefficient. The work in \cite{koda2020differentially} proposed a more energy-efficient strategy to guarantee DP by adjusting the alignment coefficient instead of adding artificial noise. The authors in \cite{liu2020privacy} studied a scheme where both the alignment coefficient and the scale of added artificial noise were adjustable. All the above works considered full device participation, where the alignment coefficient is limited by the device with the worst channel condition. Otherwise, a larger alignment coefficient will cause those devices with poor channel condition to violate the maximum transmission power constraint. Consequently, the signal-to-noise ratio (SNR) of the FL system with all devices participation will be very low and harms the learning performance. In \cite{yan2022private,cao2021optimized}, the authors proposed the misaligned OTA-FL to overcome the limited SNR issue. With misaligned OTA-FL, the gradients do not need to be aligned and are weighted by the channel condition of each device in the aggregation, therefore, the SNR will not be limited by the worst channel condition. However, it normally results in biased estimates of the averaging gradients, which will also result in a less accurate model. 

While the aforementioned literature laid a solid foundation in designing DP-OTA-FL, limited work has been done in the device scheduling to improve the alignment coefficient by selecting the devices to participate in the training. Therefore, we propose a device scheduling scheme for the aligned DP-OTA-FL, referred to as S-DPOTAFL.
%Against the above background,  to improve the alignment coefficient by selecting devices with better channel conditions to participate in the training. 
The privacy and convergence analysis are conducted to demonstrate the impact of device scheduling on privacy protection and learning performance. To minimize the impact of device scheduling on the training process, we formulate an optimization problem aiming to minimize the training loss subjecting to privacy protection and transmit power constraints. We also theoretically prove that S-DPOTAFL can perform better than the DP-OTA-FL without device scheduling (NoS-DPOTAFL). The performance of the proposed S-DPOTAFL is evaluated through simulations.
\section{System Model and preliminaries}\label{section1}

\begin{figure}[ht]
	\centering
	\includegraphics[scale=0.95]{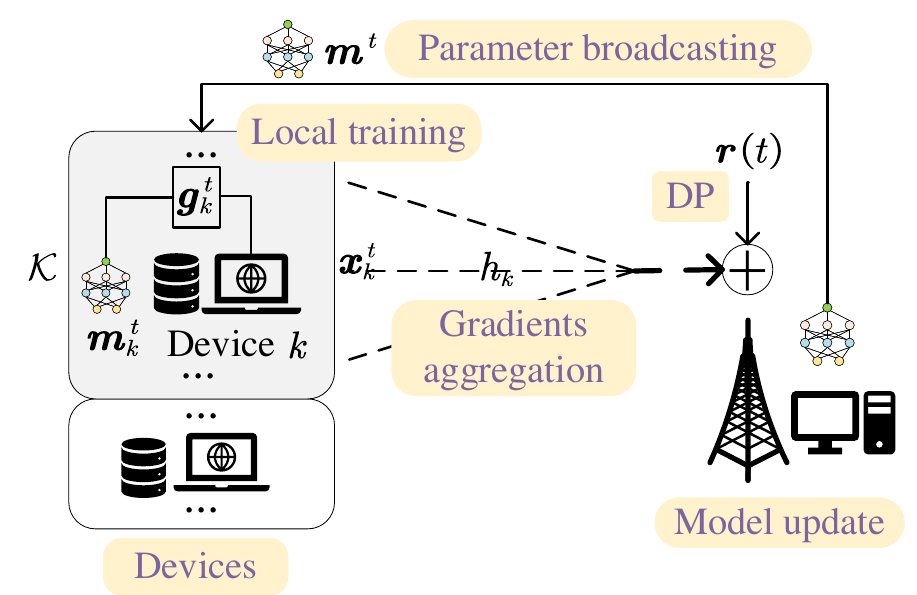}
	\caption{Differentially private OTA-FL with device scheduling.} \label{systemmodel}
\end{figure}

As shown in Fig. \ref{systemmodel}, we consider an FL system where $N$ edge devices, indexed by $\mathcal{N}$, and a BS collaboratively train a model.  Assume that the BS is curious and attempts to probe sensitive information from the received gradient. In the S-SPOTAFL framework, we select those devices whose privacy can be guaranteed by channel noise to participate in training. 

\subsection{Federated Learning}\label{federatedlearning}
 Assume that each device of index $k\in \mathcal{N}$ stores a local dataset $\mathcal{D} _k$ which contains $D_k$ pairs of training samples $\left( \boldsymbol{u},v \right)$ where $\boldsymbol{u}$ is the raw data and $v$ is the corresponding label. For simplicity, we assume that $D_1=\cdot \cdot \cdot =D_N$. Mathematically, the goal of an FL task is to obtain the model parameter that can minimize the loss function given as follows:
\begin{equation}\label{globallossfunction}
	\begin{aligned}
		\min_{\boldsymbol{m}} L\left( \boldsymbol{m} \right) =\frac{1}{N}\sum_{k=1}^N{L_k\left( \boldsymbol{m} \right)},
	\end{aligned}
\end{equation}
where $\boldsymbol{m }\in \mathbb{R} ^d$ is the model parameter to be optimized. More specifically, the objective function of device $k$ is defined as follows:
\begin{equation}\label{wg15}
	\begin{aligned}
		L_k\left( \boldsymbol{m} \right) =\frac{1}{D_k}\sum_{\left( \boldsymbol{u},v \right) \in \mathcal{D} _k}{l\left( \boldsymbol{m};\left( \boldsymbol{u},v \right) \right)},
	\end{aligned}
\end{equation}
where $l\left( \boldsymbol{m};\left( \boldsymbol{u},v \right) \right)$ is an empirical loss function defined by learning task, quantifying the loss of $\boldsymbol{m}$ at sample $\left( \boldsymbol{u},v \right)$. 

To solve the problem in (\ref{globallossfunction}), gradient descent (GD) can be applied. The main procedure of a general GD taking round $t+1$ as an example is given as follows:  \emph {Parameters broadcasting}: At the beginning of the training round, the BS broadcasts the latest global model parameter $\boldsymbol{m}^{t}$ to all the devices.  \emph{Local training}: Each device first performs the initialization of the local model by setting the received global model parameter as the local model parameter, i.e., $\boldsymbol{m}_{k}^{t} = \boldsymbol{m}^{t}, \forall k$. Then,  each device computes the gradient by
	\begin{equation}\label{wg15}
	\begin{aligned}
		\boldsymbol{g}_{k}^{t}\triangleq \nabla L_k\left( \boldsymbol{m}_{k}^{t} \right) =\frac{1}{D_k}\sum_{\left( \boldsymbol{u},v \right) \in \mathcal{D} _k}{\nabla l\left( \boldsymbol{m}_{k}^{t};\left( \boldsymbol{u},v \right) \right)},
	\end{aligned}
\end{equation}
which is sent to the BS for aggregation. \emph{Gradients aggregation}: Upon receiving all the gradients from the participants, the BS makes an aggregation of the received gradients as follows:
\begin{equation}\label{wg15}
	\begin{aligned}
		\boldsymbol{g} ^{t} =\frac{1}{N} \sum_{k=1}^{N}\boldsymbol{g} _{k}^{t} .
	\end{aligned}
\end{equation}
\emph{Model update}: The BS performs global model update as follows:
	\begin{equation}\label{globalupdate}
		\begin{aligned}
			\boldsymbol{m}^{t+1}=\boldsymbol{m}^t-\tau \boldsymbol{g}^t,
		\end{aligned}
	\end{equation}
where $\tau$ is the learning rate (also termed as step size in GD). 
The above iteration steps are repeated until a certain training termination condition is met.

\subsection{Differential Privacy}\label{differentialprivacy}
DP \cite{dwork2014algorithmic} is defined on the conception of the adjacent dataset, which guarantees the probability that any two adjacent datesets output the same result is less than a constant with the help of adding random noise. More specifically, DP quantifies information leakage in FL by measuring the sensitivity of the the gradients to the change of a single data point in the input dataset. The basic definition of $\left( \epsilon ,\xi  \right)$-DP is given as follows.
\newtheorem{definition}{Definition}
\begin{definition}$\left( \epsilon ,\xi \right)$-DP \cite{dwork2014algorithmic}:
	A randomized mechanism $\mathcal{O}$ guarantees $\left( \epsilon ,\xi  \right)$-DP if for two adjacent datasets $\mathcal{D} ,\mathcal{D} '$ differing in one sample, and measurable output space $\mathcal{Q} $ of $\mathcal{O}$, it satisfies,
	\begin{equation}\label{wg11}
		\begin{aligned}
			\mathrm{Pr}\left[ \mathcal{O} \left( \mathcal{D} \right) \in \mathcal{Q} \right] \leqslant e^{\epsilon}\mathrm{Pr}\left[ \mathcal{O} \left( \mathcal{D} ' \right) \in \mathcal{Q} \right] +\xi .
		\end{aligned}
	\end{equation} 
\end{definition}
The additive term $\xi $ allows for breaching $\epsilon$-DP with the probability $\xi $ while $\epsilon$ denotes the protection level and a smaller $\epsilon$ means a higher privacy preservation level. Specifically, the Gaussian DP mechanism which guarantees privacy by adding artificial Gaussian noise is introduced as follows.
\begin{definition}Gaussian mechanism \cite{dwork2014algorithmic}:
	A mechanism $\mathcal{O}$ is called as a Gaussian mechanism, which alters the output of another algorithm $\mathcal{L} :\mathcal{D} \rightarrow \mathcal{Q}$ by adding Gaussian noise, i.e., 
	\begin{equation}\label{wg11}
		\begin{aligned}
			\mathcal{O} \left( \mathcal{D} \right) =\mathcal{L} \left( \mathcal{D} \right) +\mathcal{N} \left( 0,\sigma ^2\mathbf{I}_d \right).
		\end{aligned}
	\end{equation} 
	Gaussian mechanism $\mathcal{O}$ guarantees $\left(\epsilon ,\xi  \right)$-DP with $\epsilon =\frac{\varDelta S}{\sigma}\sqrt{2\ln \left( \frac{1.25}{\xi } \right)}$ where $\varDelta S\triangleq \underset{\mathcal{D} ,\mathcal{D} '}{\max}\left\| \mathcal{L} \left( \mathcal{D} \right) -\mathcal{L} \left( \mathcal{D} ' \right) \right\| _2$ is the sensitivity of the algorithm $\mathcal{L}$  quantifying the sensitivity of the algorithm $\mathcal{L}$ to the change of a single data point. 
\end{definition}
%According to the Gaussian mechanism described above, privacy leakage depends both on the sensitivity of the algorithm $\mathcal{L}$ and on the power of the added Gaussian noise.

\section{DP-OTA-FL with Device Scheduling}\label{federatedlearning}
Inspired by \cite{seif2020wireless}, we consider the aligned OTA-FL. To avoid the problem that the alignment coefficient is limited by the device with the worst channel condition in the system, we propose the S-DPOTAFL. The details are described as follows.

Assume that $P_k$ is the maximum transmission power of device $k$ and $\mathcal{K}\subseteq \mathcal{N}$ is the set of the scheduled devices.  The signal sent from device $k$ in training round $t+1$ is given as:
\begin{equation}\label{inputsignal}
	\begin{aligned}
		\boldsymbol{x}_{k}^{t}=e^{-j\psi _k}\left[ \frac{\sqrt{\varphi _kP_k}}{\varpi}\boldsymbol{g}_{k}^{t} \right],\forall k\in \mathcal{K} ,
	\end{aligned}
\end{equation}
where $e^{-j\psi _k}$ is the local phase correction performed by the device $k$.
% so that the received channel coefficient is non-negative. 
$\varphi _k\in \left[ 0,1 \right] $ is the power scaling factor and we also assume $\left\| \boldsymbol{g}_{k}^{t} \right\| _2 \le \varpi$ so that $\mathrm {E}   \left [ \left \|  \boldsymbol{x} \right \| _{2}^{2}  \right ] \le P_{k} $. Consequently, the received signal at the BS is
\begin{equation}\label{receivedsignal1}
	\begin{aligned}
		\boldsymbol{y}\left( t \right) =&\sum_{k\in \mathcal{K}}{h_k\boldsymbol{x}_{k}^{t}}+\boldsymbol{r}\left( t \right) 
		\\
		=&\sum_{k\in \mathcal{K}}{\left| h_k \right|\frac{\sqrt{\varphi _kP_k}}{\varpi}\boldsymbol{g}_{k}^{t}}+\boldsymbol{r}\left( t \right),
	\end{aligned}
\end{equation}
where $h_k=\left| h_k \right|e^{j\psi _k}$ is the complex-valued time-invariant channel coefficient between device $k$ and the BS. $\boldsymbol{r}\left( t \right) \sim \mathcal{N} \left( 0,\sigma ^2\mathbf{I}_d \right) $ is the received noise at the BS, which is employed to prevent privacy leakage in this paper.

In order to obtain an unbiased estimate of the averaging gradient, all users adjust the coefficients $\varphi _k
$ to align the transmitted local gradient by the alignment coefficient $\nu$ as follows:
\begin{equation}\label{alignmentcoefficient}
	\begin{aligned}
		\left| h_k \right|\frac{\sqrt{\varphi _kP_k}}{\varpi}=\nu,\forall k\in \mathcal{K}.
	\end{aligned}
\end{equation}
It thus follows (\ref{alignmentcoefficient}) that
\begin{equation}\label{wg4}
	\begin{aligned}
		\varphi _k=\frac{\nu ^2\varpi ^2}{\left| h_k \right|^2P_k},\forall k\in \mathcal{K} .
	\end{aligned}
\end{equation}
To make sure that $\varphi _k\leqslant 1$, the following condition needs to be satisfied:
\begin{equation}\label{wg5}
	\begin{aligned}
		\nu \leqslant \frac{\underset{s\in \mathcal{K}}{\min}\left\{ \left| h_s \right|\sqrt{P_s} \right\}}{\varpi}.
	\end{aligned}
\end{equation}
%To maximize the signal power of the aligned gradient, we choose the upper bound of $\nu$ as the alignment coefficient $\nu$, i.e.,
%\begin{equation}\label{wg6}
%	\begin{aligned}
%		\nu =\frac{\underset{s\in \mathcal{K}}{\min}\left\{ \left| h_s \right|\sqrt{P_s} \right\}}{\varpi}
%	\end{aligned}
%\end{equation}
%Plugging (\ref{wg5}) back in (\ref{wg4}), we have
%\begin{equation}\label{wg6}
%	\begin{aligned}
%		\varphi _k\leqslant \frac{\underset{s\in \mathcal{K}}{\min}\left\{ \left| h_s \right|^2P_s \right\}}{\left| h_k \right|^2P_k},\forall k\in \mathcal{K} 
%	\end{aligned}
%\end{equation}
According to such aggregation scheme  described above,  the received signal at the BS  in (\ref{receivedsignal1}) can be simplified as:
\begin{equation}\label{receivedsignal2}
	\begin{aligned}
		\boldsymbol{y}\left( t \right) =\nu \sum_{k\in \mathcal{K}}{\boldsymbol{g}_{k}^{t}}+\boldsymbol{r}\left( t \right) .
	\end{aligned}
\end{equation}
It can be learned that a smaller $\nu$ results in a low SNR of the FL system. 
In order to estimate the averaging gradient, the BS performs post-processing as follows:
\begin{equation}\label{gradientestimate}
	\begin{aligned}
		\boldsymbol{\tilde{g}}^t=\frac{1}{\left| \mathcal{K} \right|\nu}\boldsymbol{y}\left( t \right) 
		=\frac{1}{\left| \mathcal{K} \right|}\sum_{k\in \mathcal{K}}{\boldsymbol{g}_{k}^{t}}+\frac{1}{\left| \mathcal{K} \right|\nu }\boldsymbol{r}\left( t \right), %=\boldsymbol{g}^t+\frac{1}{\left| \mathcal{K} \right|\nu }\boldsymbol{r}\left( t \right) 
	\end{aligned}
\end{equation}
which is finally used to update global model as follows:
\begin{equation}\label{gradientestimate}
	\begin{aligned}
		\boldsymbol{m}^{t+1}=\boldsymbol{m}^t-\tau  \boldsymbol{\tilde{g }} ^t.
	\end{aligned}
\end{equation}

\section{Privacy and Convergence Analysis and Problem Formulation}\label{section4}

To illustrate the impact of device scheduling on privacy protection and learning performance,  we conduct privacy and convergence analysis. Then, based on these analytical results, we formulate an optimization problem to minimize the optimality gap with consideration of privacy and transmit power constraints.
\subsection{Privacy and Convergence Analysis}

\subsubsection{Assumptions}
For analysis purposes, we provide the following assumptions first.
%[Smoothness]
\newtheorem{assumption}{Assumption}
\begin{assumption}\label{gradientassumption}
	The expected squared norm of each gradient is bounded:
	\begin{equation}\label{wg19}	
		\begin{aligned}
			\mathbb{E} \left[ \left\| \boldsymbol{g}_{k}^{t} \right\| _2 \right] \leqslant \varpi .
		\end{aligned}
	\end{equation}
\end{assumption}

\begin{assumption}\label{assumption1}
	Assume that $L\left( \cdot \right) \,\,$ is $\zeta$-smooth, i.e., for all $\boldsymbol{\iota }'$ and $\boldsymbol{\iota }$, one has
	\begin{equation}\label{wg8}
		\begin{aligned}
			L\left( \boldsymbol{\iota }' \right) -L\left( \boldsymbol{\iota } \right) \leqslant \left( \boldsymbol{\iota }'-\boldsymbol{\iota } \right) ^{\mathrm{T}}\nabla L\left( \boldsymbol{\iota } \right) +\frac{\zeta}{2}\left\| \boldsymbol{\iota }'-\boldsymbol{\iota } \right\| _{2}^{2}.
		\end{aligned}
	\end{equation}
\end{assumption}

\begin{assumption}\label{Polyak}
	Assume that $L\left( \cdot \right)$ satisfies Polyak-Lojasiewicz inequality,  i.e., for all $\boldsymbol{\iota } $, there is a constant $\varrho \geqslant 0$ satisfying,  
	\begin{equation}\label{wg20}
		\begin{aligned}
			\left\| \nabla L\left( \boldsymbol{\iota } \right) \right\| _{2}^{2}\geqslant 2\varrho\left( L\left( \boldsymbol{\iota } \right) -L\left( \boldsymbol{\iota }^* \right) \right).
		\end{aligned}
	\end{equation}
	%where $\boldsymbol{\theta }^*$ denotes the optimal model.
\end{assumption}

\subsubsection{Privacy analysis}
We here present the privacy analysis of the S-DPOTAFL.
\begin{lemma}\label{lemma2}
	Assume that Assumption 1 holds. S-DPOTAFL guarantees $\left( \epsilon _k,\xi  \right)$-DP of device $k\in\mathcal{K}$ where
	\begin{equation}\label{privacyleakage}
		\begin{aligned}
			\epsilon _k=\frac{2\varpi \nu }{\sigma}\cdot \phi,% \leqslant \frac{2\underset{s\in \mathcal{K}}{\min}\left\{ \left| h_s \right|\sqrt{P_s} \right\}}{\sigma}\cdot \phi ,
		\end{aligned}
	\end{equation}
	where $\phi =\sqrt{2\ln \frac{1.25}{\xi }}$.
\end{lemma} 

\itshape {Proof:}  \upshape Here we use index $m$ instead of $k$ to avoid confusion between the specific index of device $k$ and the notation $k$ in the summation. 
Assume that $\mathcal{D} _m$ and $\mathcal{D} _{m}^{'}$ are two adjacent datasets differing in one sample. $\boldsymbol{y}'\left(t\right) $ is the received signal at the BS, which only differs in one gradient with $ \boldsymbol{y}\left(t\right)$. The gradient $\left( \boldsymbol{g}_{m}^{t} \right) '$ from device $m$ in $\boldsymbol{y}'\left(t\right) $ is obtained based on $\mathcal{D} _{m}^{'}$.
Based on the definition of sensitivity and Assumption \ref{gradientassumption}, one has
\begin{equation}\label{wg15}
	\begin{aligned}
		\varDelta S_{m}\triangleq& \underset{\mathcal{D} _m,\mathcal{D} _{m}^{'}}{\max}\left\| \boldsymbol{y}\left(t\right)-\boldsymbol{y}'\left(t\right) \right\| _2
		=\underset{\mathcal{D} _m,\mathcal{D} _{m}^{'}}{\max}\left\|\nu \left( \boldsymbol{g}_{m}^{t}-\left( \boldsymbol{g}_{m}^{t} \right) ' \right) \right\| _2
		\\
		=&\nu \left\| \boldsymbol{g}_{m}^{t}-\left( \boldsymbol{g}_{m}^{t} \right) ' \right\| _2	\overset{\left( a \right)}{\leqslant}2\varpi\nu ,
	\end{aligned}
\end{equation}
where (a) is from triangular inequality and Assumption \ref{gradientassumption}.
According to Definition 2 and the above result, one completes the proof of Lemma \ref{lemma2} by replacing $m$ with $k$.
$\hfill \blacksquare$
%\textit{Proof:} : Please refer to Appendix \ref{proofofprivacyanalysis}. 

Lemma \ref{lemma2} shows the impact of the alignment coefficient on privacy leakage. More specifically, a smaller alignment coefficient $\nu$ leads to less privacy leakage.

\begin{remark}
	Note that when the ``$=$" in (\ref{privacyleakage}) is replaced by ``$\leqslant$", it indicates a stronger privacy protection so it still satisfies $\left( \epsilon _{k},\xi  \right)$-DP.
\end{remark}

\subsubsection{Convergence analysis}\label{convergenceanalysis}
We here present the results of convergence analysis to show the impact of devices scheduling and alignment coefficient on the training process. Assume that the training process terminates after $T$ rounds and $\boldsymbol{m}^T $ is the obtained model and $\boldsymbol{m}^* $ is the optimal global model. Then, we have the following results.
\begin{theorem}\label{lemmaconvergenceanalysis}
 Given the learning rate $\tau =\frac{1}{\zeta}$, the upper bound of the optimality gap $\mathbb{E} \left[ L\left( \boldsymbol{m}^T \right) -L\left( \boldsymbol{m}^* \right) \right]$ is given by
	\begin{equation}\label{optimalitygap1}
		\begin{aligned}
			&\mathbb{E} \left[ L\left( \boldsymbol{m}^T \right) -L\left( \boldsymbol{m}^* \right) \right] 
			\\
			\leqslant& \eta ^{T-1}\mathbb{E} \left[ L\left( \boldsymbol{m}^0 \right) -L\left( \boldsymbol{m}^* \right) \right] +\frac{\varPhi \left( \mathcal{K},\nu  \right) }{2\zeta}\frac{1-\eta ^{T-1}}{1-\eta},
		\end{aligned}
	\end{equation}
	where 
	\begin{equation}\label{convercoeff}
		\begin{aligned}
			\eta =1-\frac{\varrho}{\zeta},
		\end{aligned}
	\end{equation} and
	\begin{equation}\label{deviceselectionimpact}
		\begin{aligned}
			\varPhi \left( \mathcal{K},\nu  \right) =4\varpi ^2\left( 1-\frac{\left| \mathcal{K} \right|}{N} \right) ^2+\frac{d\sigma ^2}{\left| \mathcal{K} \right|^2\nu ^2},
		\end{aligned}
	\end{equation}
	which characterizes the impact of alignment coefficient and device scheduling on the optimality gap. 
	The expectation is with respect to the randomness of Gaussian noise.
\end{theorem} 
	\textit{Proof:} Please refer to Appendix \ref{proofofConvergence}.$\hfill \blacksquare$ 

The first term on the right-hand side of the optimality gap is the initial gap which decreases with $T$ as $\eta \le 1$.  The second term reveals the impact of alignment coefficient and device scheduling on learning performance. More specifically, the larger number of scheduled devices and alignment coefficient benefit the training process.  It can be understood by observing (\ref{receivedsignal2}) that the channel noise makes smaller distortion to the gradients when more devices are involved. Additionally, a larger alignment coefficient means a higher SNR of the system. From (\ref{wg5}), we can learn that the alignment coefficient $\nu$ is limited by the scheduled device with the worst channel condition, i.e., $\underset{s\in \mathcal{K}}{\min}\left\{ \left| h_s \right|^2P_s \right\}$. Therefore, an appropriate device selection is significant for improving learning performance while preserving privacy.

Based on Theorem 1, we can also derive the optimality gap of an FL system with full device participation and a noise-free channel.
\begin{corollary} \label{corollaryoptimalitygap}
	Given the learning rate $\tau =\frac{1}{\zeta}$, the upper bound of the optimality gap $\mathbb{E} \left[ L\left(\boldsymbol{m}^T \right) -L\left( \boldsymbol{m}^* \right) \right]$ of the FL algorithm without considering noise and device scheduling is
	\begin{equation}\label{goptimalitygap2}
		\begin{aligned}
			&\mathbb{E} \left[ L\left( \boldsymbol{m}^T \right) -L\left( \boldsymbol{m}^* \right) \right] 
			\\
			\leqslant& \left( 1-\frac{\varrho}{\zeta} \right) ^{T-1}\mathbb{E} \left[ L\left( \boldsymbol{m}^0 \right) -L\left( \boldsymbol{m}^* \right) \right] .
		\end{aligned}
	\end{equation}
	%		where $\chi =\underset{t}{\max}\left\{ 2\frac{\varGamma}{\tau ^0}+\vartheta ^2+G^2\varPsi ^t \right\}.$
\end{corollary}
\itshape {Proof:}  \upshape   Since the FL algorithms do not consider the noise and device selection, we have $\sigma =0
$ and $\left| \mathcal{K} \right|=N$, Hence, $4\varpi ^2\left( 1-\frac{\left| \mathcal{K} \right|}{N} \right) ^2+\frac{d\sigma ^2}{\left| \mathcal{K} \right|^2\nu^2}=0$. Then (\ref{goptimalitygap2}) can be derived based on (\ref{optimalitygap1}). 	
\hfill $\blacksquare$

From Corollary \ref{corollaryoptimalitygap}, we can observe that, if we do not consider the noise, the FL algorithm with full device participation will converge to the optimal global FL model without any gaps. This result corresponds to the results in the existing works \cite{chen2020joint,friedlander2012hybrid}. 

\subsection{Optimization Problem for Device Scheduling}
In this section, our goal is to minimize the optimality gap via the selection of participants and alignment coefficient considering privacy preservation. Assume that each device has the same requirement of privacy protection, i.e., $\left( \epsilon,\xi  \right)$. By omitting the terms that are irrelevant to device scheduling and alignment coefficient, the optimization problem is formulated as follows:
\begin{align}\label{P1}
	\mathbf{P}1.&\quad \underset{\mathcal{K} ,\nu}{\min}\left\{ 4\varpi ^2\left( 1-\frac{\left| \mathcal{K} \right|}{N} \right) ^2+\frac{d\sigma ^2}{\left| \mathcal{K} \right|^2\nu ^2} \right\} 
	\\
	\mathbf{s}.\mathbf{t}.&\quad \mathcal{K} \subseteq \mathcal{N} ,
	\tag{25a}
	\\
	& \quad \nu \leqslant \frac{\underset{s\in \mathcal{K}}{\min}\left\{ \left| h_s \right|\sqrt{P_s} \right\}}{\varpi}, \tag{25b}
	\\
	& \quad \frac{2\varpi \nu }{\sigma}\cdot \phi \leqslant \epsilon. \tag{25c}
\end{align}
Constraint (25b) implies that the alignment coefficient should ensures that $\varphi _k\leqslant 1$ as mentioned in (\ref{wg5}). Constraint (25c) is privacy requirement. By defining $\theta =\varpi \nu $ and $c_{\left[ \mathcal{K} \right]}=\underset{s\in \mathcal{K}}{\min}\left\{ \left| h_s \right|\sqrt{P_s} \right\}$, Problem $\mathbf{P}1$ can be equivalently rewritten as:
\begin{align}\label{wg33}
	\mathbf{P}2.& \quad \underset{\mathcal{K} ,\theta}{\min}\left\{ 4\left( 1-\frac{\left| \mathcal{K} \right|}{N} \right) ^2+\frac{d\sigma ^2}{\left| \mathcal{K} \right|^2\theta ^2} \right\} 
	\\
	\mathbf{s}.\mathbf{t}.&\quad \mathcal{K} \subseteq \mathcal{N}, 
	\tag{26a}
	\\
	& \quad \theta \leqslant \min \left\{ c_{\left[ \mathcal{K} \right]},\frac{\epsilon \sigma}{2\phi} \right\} .\tag{26b}
\end{align}
Assume that the elements in $\boldsymbol{c}=\left[ c_1,...c_i,...c_N \right]$ where $c_i=\left| h_i \right|\sqrt{P_i}$ are sorted in ascending order. By observering the objective function, we learn that larger $\left| \mathcal{K} \right|$ and $\theta$ yields a better objective function value. However, $\theta$ is equivalent to the threshold at which the device is qualified. As $\theta$ increases, the value of $\left| \mathcal{K} \right|$ decreases.  Conversely, a larger $\left| \mathcal{K} \right|$ leads to a smaller $\theta$. Therefore, there is a tradeoff between $\theta$ and $\left| \mathcal{K} \right|$. 

For the clarity, we give the solution in two cases: 1) $c_1>\frac{\epsilon \sigma}{2\phi}$; 2) $c_1\leqslant \frac{\epsilon \sigma}{2\phi}$ as follows:

\subsubsection{In the case that $c_1> \frac{\epsilon \sigma}{2\phi}$} Constraint (26b) can be rewritten as $ \theta \leqslant  \frac{\epsilon \sigma}{2\phi} $. Then, we have the optimal solution as follows.
\begin{lemma}\label{solution1}
	The optimal solution to $\mathbf{P}2$ when $c_1> \frac{\epsilon \sigma}{2\phi}$ is
	\begin{align}\label{wg37}
		\theta =\frac{\epsilon \sigma}{2\phi},\quad \mathcal{K} =\mathcal{N},
	\end{align}	 	
in which case S-SPOTAFL is equivalent to NoS-DPOTAFL.
\end{lemma} 
\itshape {Proof:}  \upshape Firstly, to achieve a larger $\theta$, we have $\theta =\frac{\epsilon \sigma}{2\phi}$. On the other hand, all the $k$ with $c_k\ge \frac{\epsilon \sigma}{2\phi}$ should be selected to achieve a larger $\left| \mathcal{K} \right|$, i.e., a better value of objective function.  Since $\frac{\epsilon \sigma}{2\phi}<  c_1\le c_k,\forall k\in \mathcal{N} $, we have $\mathcal{K} =\mathcal{N}$. This completes the proof of Lemma \ref{solution1}. 
\hfill $\blacksquare$

\subsubsection{In the case that $c_1\leqslant \frac{\epsilon \sigma}{2\phi}$}
\begin{figure}[ht]
	\centering
	\includegraphics[scale=0.9]{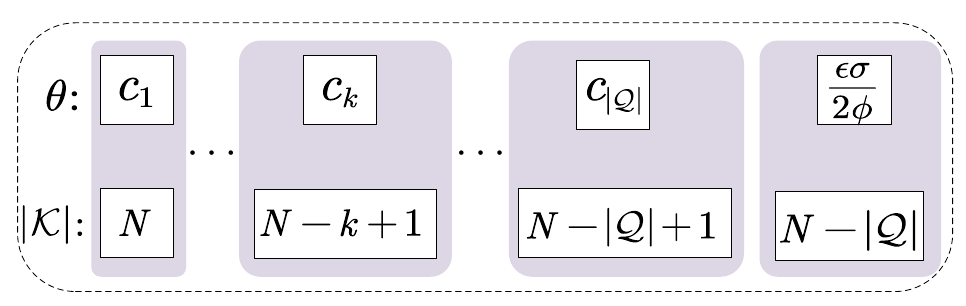}
	\caption{The relationship between the potential optimal $\left| \mathcal{K} \right|$ and $\theta$.} \label{figrelationship}
\end{figure}

, we define 
\begin{align}\label{wg37}
	\mathcal{Q} =\left\{ \left. k \right|c_1\leqslant c_k< \frac{\epsilon \sigma}{2\phi} \right\}.
\end{align}
\addtolength{\topmargin}{0.06in}
Then, we have $c_{\left | \mathcal{Q}  \right | }< \frac{\epsilon \sigma}{2\phi} \le c_{\left | \mathcal{Q}\right|+1}$. Constraint (26b) can be discussed in two cases: (1) $\theta \le c_{\left [ \mathcal{K}  \right ] }=c_k,k\in\mathcal{Q} $; (2) $\theta \le \frac{\epsilon \sigma}{2\phi}$. Then, we have the following results.
\begin{lemma}\label{relationship}
	The minimum value of $\left| \mathcal{K} \right|$ is $N-\left| \mathcal{Q} \right|$. The relationship between the potential optimal solution pairs, i.e.,  $\left| \mathcal{K} \right|$ and $\theta$, as shown in Fig. \ref{figrelationship} can be given by
	\begin{align}\label{wg37}
		\theta=
		\begin{cases}c_{N-\left| \mathcal{K} \right|+1},if \left| \mathcal{K} \right| \ge N-\left| \mathcal{Q} \right|+1
			\\\frac{\epsilon \sigma}{2\phi}, if \left| \mathcal{K} \right| = N-\left| \mathcal{Q} \right|
		\end{cases}.
	\end{align}
\end{lemma} 
\itshape {Proof:}  \upshape 
Firstly, given a value of $\left| \mathcal{K} \right| \ge N-\left| \mathcal{Q} \right|+1$, the largest $c_{\left[ \mathcal{K} \right] }$ that can be achieved is $c_{N-\left| \mathcal{K} \right|+1}$. To achiebe a larger $\theta$, we have $\theta=c_{N-\left| \mathcal{K} \right|+1}$. The largest feasible value of $\theta$ is $\frac{\epsilon \sigma}{2\phi}$, at which $\left| \mathcal{K} \right|$ achieves the minimum value $N-\left| \mathcal{Q} \right|$. Then, we complete the proof of Lemma \ref{relationship}.
\vspace{-10pt}
%Since the value of $\left| \mathcal{K} \right|$ decrease with $c_{\left[ \mathcal{K} \right]}$, the largest $c_{\left[ \mathcal{K} \right]} $ corresponds to the minimum value of $\left| \mathcal{K} \right|$ $c_{\left| \mathcal{Q} \right|}$. The largest $c_{\left[ \mathcal{K} \right]}$ can be set is $c_{\left| \mathcal{Q} \right|}$ considering the privacy constraint, then all the $k$ with $c_k\geqslant c_{\left| \mathcal{Q} \right|}$ should be selected to yeild a better objective function. Therefore, the minimum value of $\left| \mathcal{K} \right|$ is $N-\left| \mathcal{Q} \right|+1$.
\hfill $\blacksquare$
%\itshape {Proof:}  \upshape Assume that $\theta _{\left[ \mathcal{K} \right]}=c_i,i\in \mathcal{Q} $, it is clear that a larger $\left| \mathcal{K} \right|$ yields a better objective value. Consequently, all the device $k$ with $c_k\geqslant c_i$ should be seleced, i.e., $\left| \mathcal{K} \right|=N-i+1$. On the other hand, for a given $\left| \mathcal{K} \right|\geqslant N-\left| \mathcal{Q} \right|+1$, $\theta _{\left[ \mathcal{K} \right]}$ should be set to $c_{N-\left| \mathcal{K} \right|+1}$ to achieve a larger $\theta _{\left[ \mathcal{K} \right]}$.s\hfill $\blacksquare$

Lemma \ref{relationship} offers an important insight that each $\theta$ corresponds to a $\left| \mathcal{K} \right|$. 
Then, the potential solutions to $\mathbf{P}2$ can be obtained as follows.
\begin{lemma}\label{closedsolution}
	There are $\left| \mathcal{Q} \right|+1$ closed-form solutions which may be the globally optimal solution. The $i$-th, $1\leq i \leq \left| \mathcal{Q} \right|$, possible solution $\theta_i$ and $\mathcal{K} _i$ and is given by
	\begin{align}\label{wg37}
		\theta_i = c_i, \quad \mathcal{K} _i=\left\{ \left. k \right|c_k\ge c_i \right\},
	\end{align}
and the possible solution $\theta_{\left| \mathcal{Q} \right|+1}$, $\mathcal{K} _{\left| \mathcal{Q} \right|+1}$ is 
     \begin{align}\label{wg37}
     	\theta_ {\left| \mathcal{Q} \right|+1}= \frac{\epsilon \sigma}{2\phi}, \quad \mathcal{K} _{\left| \mathcal{Q} \right|+1}=\left\{ \left. k \right|c_k \ge \frac{\epsilon \sigma}{2\phi}\right\}.
     \end{align}
\end{lemma}

\itshape {Proof:}  \upshape Firstly, there are $\left| \mathcal{Q} \right|$ elements in $\mathcal{Q}$, which are the potential value of $c_{\left[ \mathcal{K} \right]}$, i.e., $\theta$. It thus following Lemma \ref{relationship} that there are $\left| \mathcal{Q} \right|$ pairs of $\theta$ and $\left| \mathcal{K} \right|$, i.e., $\left| \mathcal{Q} \right|$ potential optimal solutions, which may achieve the best performance. Specifically, the $i$-th solution corresponds to the setting that $\theta_i=c_i$ and $\left| \mathcal{K} \right|=N-i+1$. In this case, $\mathcal{K} _i=\left\{ i,...,N \right\} $, i.e., $\mathcal{K} _i=\left\{ \left. k \right|c_k\ge c_i \right\} $. Additionally, $\theta =\frac{\epsilon \sigma}{2\phi}$ is the $\left| \mathcal{Q} \right|+1$-th solution, in which $\mathcal{K} _{\left| \mathcal{Q} \right|+1}=\left\{ \left. k \right|c_k \ge \frac{\epsilon \sigma}{2\phi}\right\}$. Then, we complete the proof of Lemma \ref{closedsolution}.
\hfill $\blacksquare$

	Based on Lemma \ref{closedsolution}, we can perform the one-dimension search method to obtain the optimal solution. The optimal solution to Problem P2 is $\mathcal{K} ^*, \theta^*$ where
\begin{align}
	\mathcal{K} ^*, \theta^*=arg\underset{1\leqslant i\leqslant \left| \mathcal{Q} \right|+1}{\min}\left\{ \varPsi \left( \mathcal{K} _i,\theta _i \right) \right\} ,
\end{align}
where $\varPsi \left( \mathcal{K} _i,\theta _i \right) =4\left( 1-\frac{\left| \mathcal{K} _i \right|}{N} \right) ^2+\frac{d\sigma ^2}{\left| \mathcal{K} _i \right|^2\theta _{i}^{2}}$.

We next present the situation that S-DPOTA-FL performs better than NoS-DPOTAFL. Since S-DPOTA-FL is equivalent to NoS-DPOTAFL when $c_1> \frac{\epsilon \sigma}{2\phi}$, we only consider the case that $c_1\leqslant \frac{\epsilon \sigma}{2\phi}$.

\begin{lemma}\label{condition}
	Assume that $c_1\leqslant \frac{\epsilon \sigma}{2\phi}$. S-DPOTAFL performs better than NoS-DPOTAFL when the following condition is satisfied:
	\begin{align}\label{wg37}
		\left| \mathcal{K} \right|\theta\geqslant \frac{1}{\sqrt{\frac{1}{N^2c_{1}^{2}}-\frac{4}{d\sigma ^2}}}.
	\end{align}
\end{lemma}

\itshape {Proof:}  \upshape NoS-DPOTAFL is equivalent to the solution that $\theta = c_1$ and $\mathcal{K} = \mathcal{N}$, in which case, the value of objective function is $\frac{d\sigma^2}{N^{2}c_1^2} $. 
%The minimum valueof $\left| \mathcal{K} \right|$ is $N-\left| \mathcal{Q} \right|+1$. 
By solving $4+\frac{d\sigma ^2}{\left| \mathcal{K} \right|^2\theta ^2}\le \frac{d\sigma^2}{N^{2}c_1^2} $, we complete the proof of Lemma \ref{condition}.
\hfill $\blacksquare$

\section{Simulation Results}\label{section5}
We evaluate the effectiveness of the S-DPOTAFL by training a convolutional neural network (CNN) on the popular MNIST dataset. In particular, CNN consists of two $5\times5$ convolution layers with the rectified linear unit (ReLU) activation. The two convolution layers have 10 and 20 channels respectively, and each layer has $2\times2$ max pooling, a fully-connected layer with 50 units and ReLU activation, and a log-softmax output layer, in which case $d= 21840$. The learning rate is set to $\eta=0.1$. 
%The MNIST dataset consists of 60,000 images for training and 10,000 testing images of the 10 digits.We have the general assumption that there is an equal number of training data samples for each device and no overlap between the local training data sets \cite{zhu2019broadband} \cite{wang2019adaptive}.We assume that local datasets are IID, where the initial training dataset is randomly divided into $N$ batches and each device is assigned to one batch. 
We assume that each device has the same transmit power $P$ and the minimal channel gain $\left|h_k\right|$ is set to $0.1$. The privacy level is set to $(\epsilon ,\xi )=(10,0.1)$ and $\sigma=1$.

\begin{figure*}[ht]
	\centering
	\subfigure[$P=5$]	
	{  \begin{minipage}[t]{0.315\linewidth}
			\centering
			\includegraphics[scale=0.6]{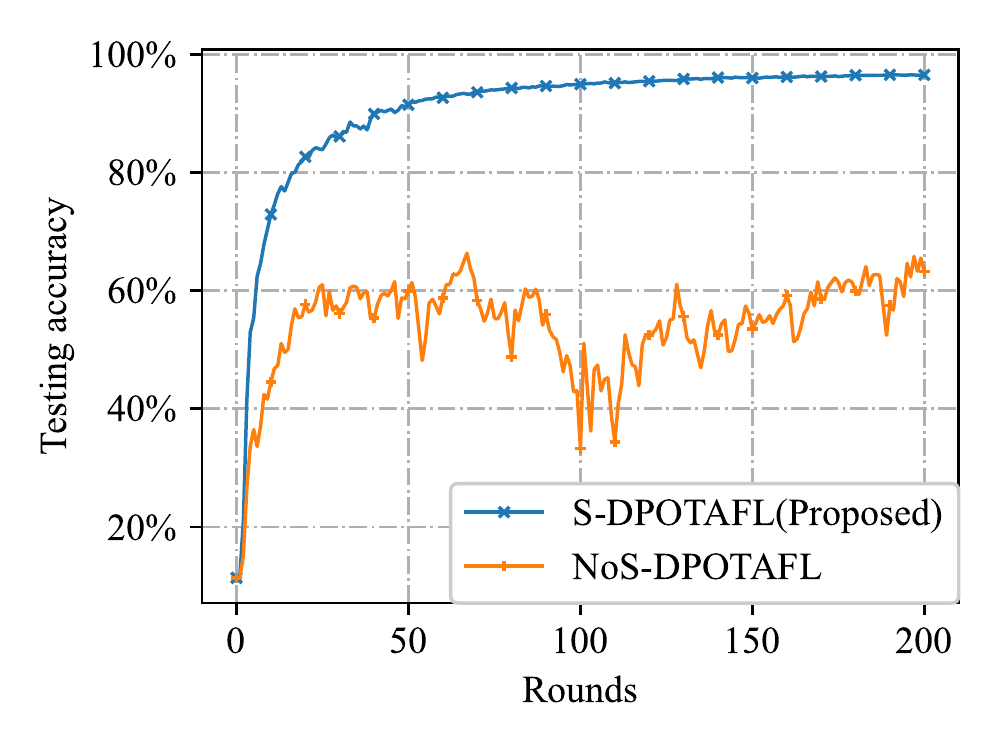}
		\end{minipage}	
	}
	\subfigure[$P=50$]	
	{  \begin{minipage}[t]{0.315\linewidth}
			\centering
			\includegraphics[scale=0.6]{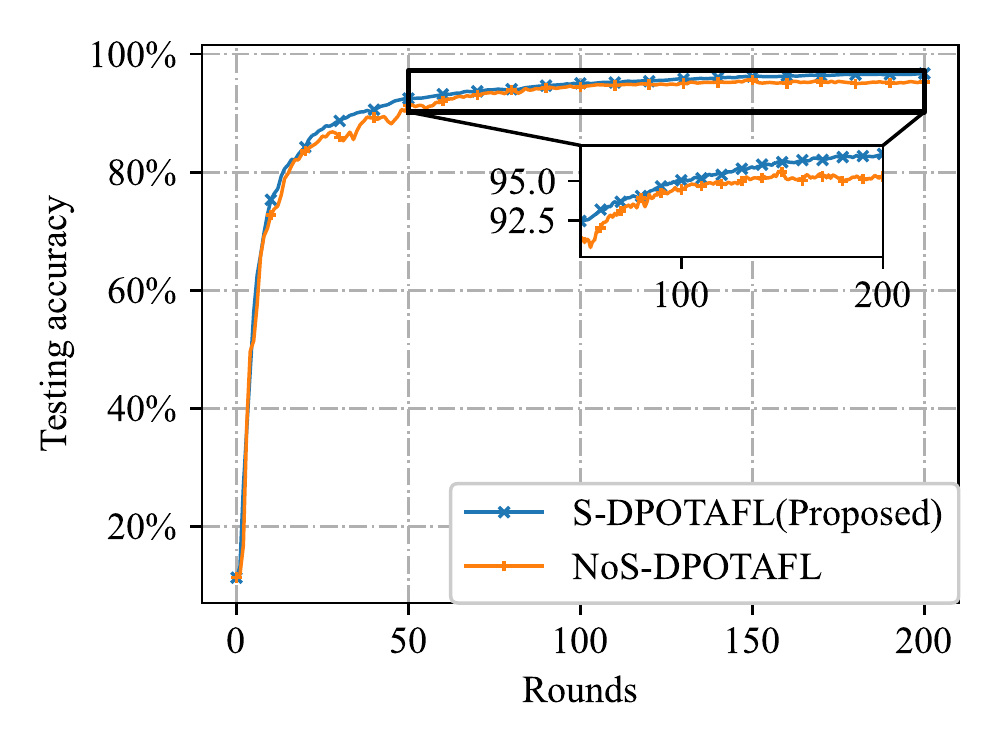}
		\end{minipage}	
	}
	\subfigure[$P=200$]	
	{  \begin{minipage}[t]{0.315\linewidth}
			\centering
			\includegraphics[scale=0.6]{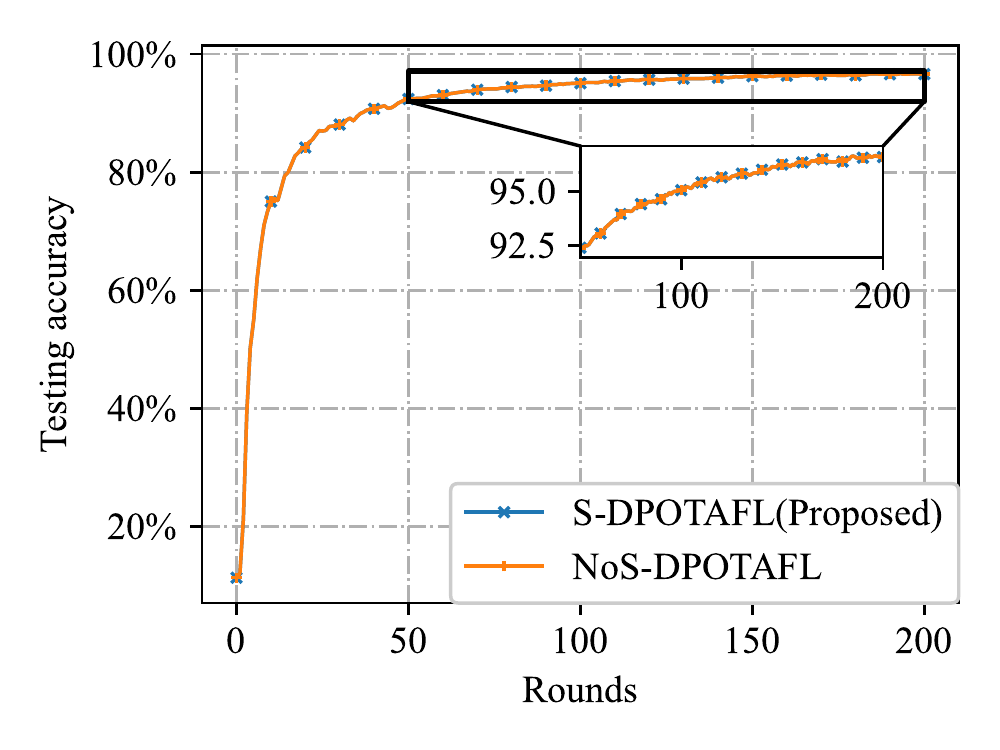}
		\end{minipage}	
	}
	\caption{Testing accuracy of the S-DPOTAFL and NoS-DPOTAFL with different $P$.} \label{testingAccuracyP}
\end{figure*}

In Fig. \ref{testingAccuracyP}, we plot the testing accuracy of the S-DPOTAFL and NoS-DPOTAFL with different $P$ where $N=50$. It can be observed that the superiority of the S-DPOTAFL is significant when $P$ is relatively smaller. In the cases that $P$ is small, although all the devices are involved in training in the NoS-DPOTAFL, the alignment coefficient in the NoS-DPOTAFL is very small, which results in a quite low SNR of the FL system. Additionally, we can learn that the performance gap between NoS-DPOTAFL and S-DPOTAFL decreases as $P$ increases. In particular, the S-DPOTAFL achieves the same accuracy as the NoS-DPOTAFL with $P=200$. This is because when $P=200$, we have $c_1> \frac{\epsilon \sigma}{2\phi}$, in which case the optimal solution to the S-DPOTAFL is equivalent to the NoS-DPOTAFL as shown in Lemma \ref{solution1}. 

\begin{figure*}[ht]\label{N}
	\centering
	\subfigure[$N=25$]	
	{  \begin{minipage}[t]{0.315\linewidth}
			\centering
			\includegraphics[scale=0.6]{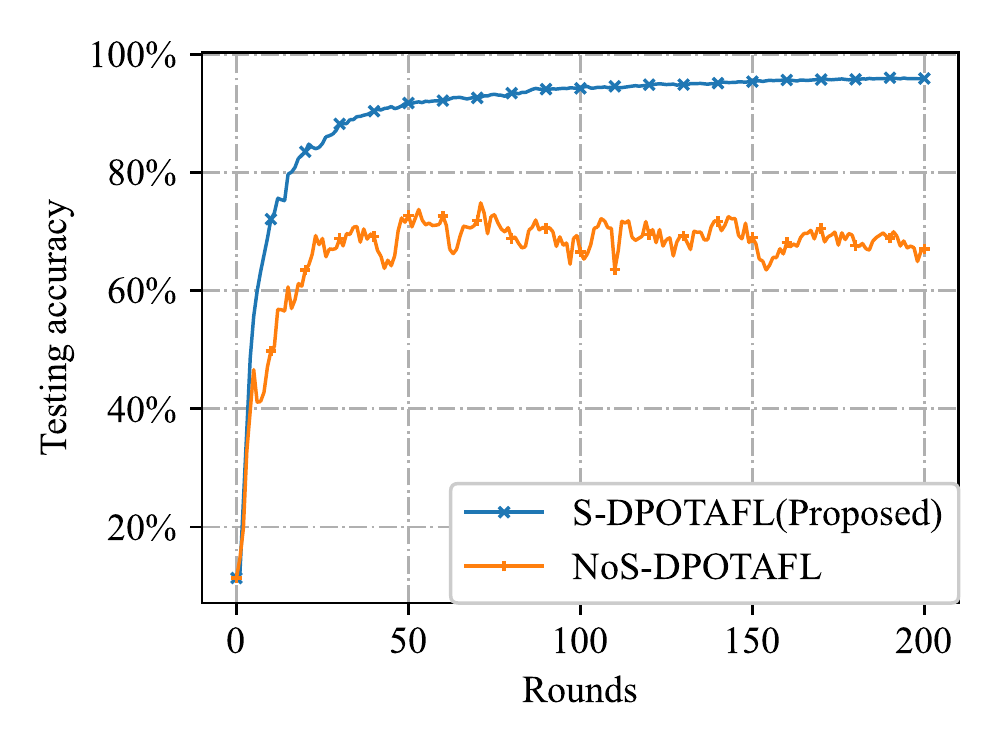}
		\end{minipage}	
	}
	\subfigure[$N=50$]	
	{  \begin{minipage}[t]{0.315\linewidth}
			\centering
			\includegraphics[scale=0.6]{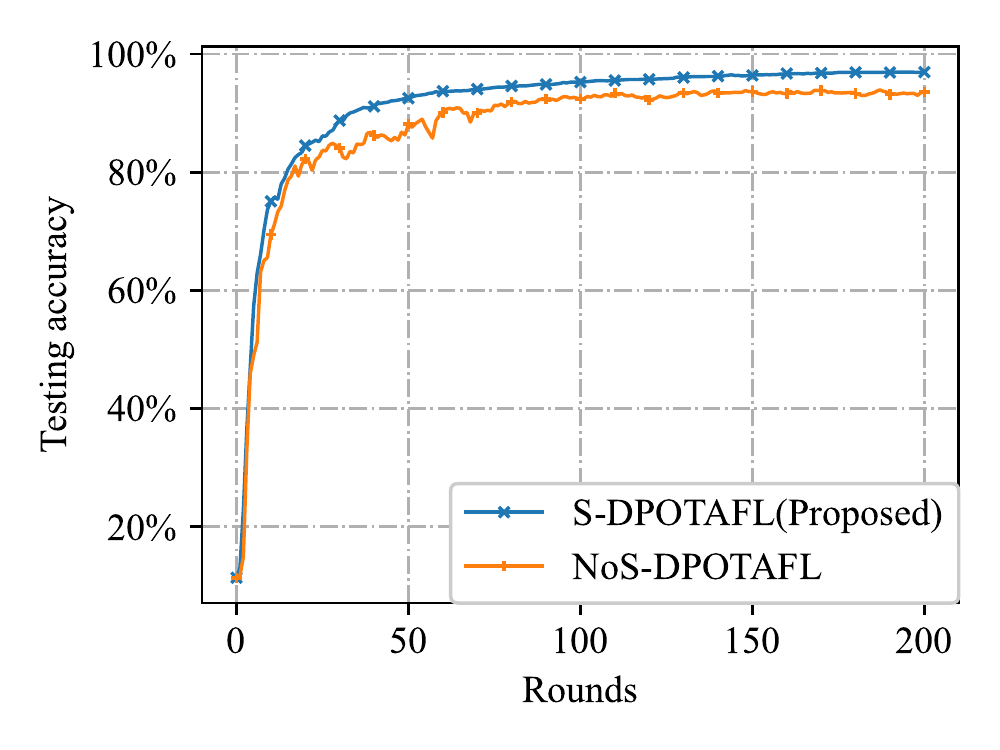}
		\end{minipage}	
	}
	\subfigure[$N=100$]	
	{  \begin{minipage}[t]{0.315\linewidth}
			\centering
			\includegraphics[scale=0.6]{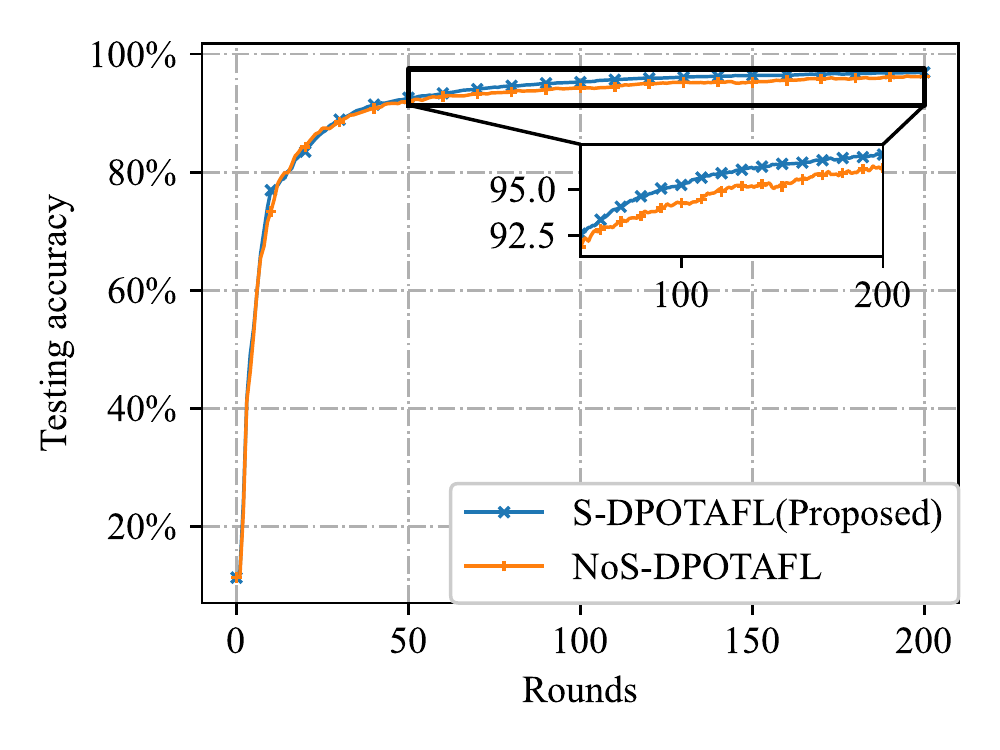}
		\end{minipage}	
	}
	\caption{Testing accuracy of the S-DPOTAFL and NoS-DPOTAFL with different $N$.} \label{testingaccuracyN}
\end{figure*}

Fig. \ref{testingaccuracyN} plots the testing accuracy of the S-DPOTAFL and NoS-DPOTAFL with different $N$ where $P=25$. The S-DPOTAFL performs better than the NoS-DPOTAFL in all the cases. The superiority of  the S-DPOTAFL is particularly noticeable in the case that $N=25$ because both the number of participants and the alignment coefficient are relatively small for the NoS-DPOTAFL, while the alignment coefficient in the S-DPOTAFL is larger by selecting the devices with better channel conditions. The performance of the NoS-DPOTADL is improved as $N$ grows because including more devices means less noise distortion to the aggregated gradients given a fixed $c_1$. 

From all the results above, we conclude that with the small-scale FL systems where the number and the power of devices are small, the S-DPOTAFL is more useful than the NoS-DPOTAFL.

\section{Conclusion}\label{section6}
The device scheduling problem for the aligned DP-OTA-FL system has been studied in this work. The privacy and convergence analysis are conducted. We have formulated an optimization problem to minimize the optimality gap considering privacy protection. The closed-form solution has been derived and we have also obtained the scenarios that the S-DPOTAFL performs better than the NoS-DPOTAFL. 

\appendices
\section{Proof of Convergence analysis}\label{proofofConvergence}
Recalling that 
\begin{equation}\label{wg15}
	\begin{aligned}
		\boldsymbol{m}^{t+1}=\boldsymbol{m}^t-\tau \left( \nabla L\left( \boldsymbol{m}^t \right) +\boldsymbol{e}^{t} \right),
	\end{aligned}
\end{equation}
where 
\begin{equation}\label{wg15}
	\begin{aligned}
		\boldsymbol{e}^{t}=\frac{1}{\left| \mathcal{K} \right|}\sum_{k\in \mathcal{K}}{\boldsymbol{g}_{k}^{t}}+\frac{1}{\left| \mathcal{K} \right|\nu }\boldsymbol{r}\left( t \right) -\nabla L\left( \boldsymbol{m}^t \right),
	\end{aligned}
\end{equation}
It thus following that
\begin{equation}\label{wg53}
	\begin{aligned}
		&\mathbb{E} \left[ L\left( \boldsymbol{m}^{t+1} \right) \right] -\mathbb{E} \left[ L\left( \boldsymbol{m}^t \right) \right] 
		\\
		\overset{\left( a \right)}{\leqslant}&-\tau \mathbb{E} \left[ \left< \nabla L\left( \boldsymbol{m}^t \right) ,\nabla L\left( \boldsymbol{m}^t \right) +\boldsymbol{e}^{t} \right> \right] 
		\\
		&+\frac{\tau ^2\zeta}{2}\mathbb{E} \left[ \left\| \nabla L\left( \boldsymbol{m}^t \right) +\boldsymbol{e}^{t} \right\| _{2}^{2} \right] 
		\\
		=&-\tau \left< \nabla L\left( \boldsymbol{m}^t \right) ,\nabla L\left( \boldsymbol{m}^t \right) \right> -\tau \left< \nabla L\left( \boldsymbol{m}^t \right) ,\mathbb{E} \left[ \boldsymbol{e}^{t} \right] \right> 
		\\
		&+\frac{\tau ^2\zeta}{2}\left\| \nabla L\left( \boldsymbol{m}^t \right) \right\| _{2}^{2}+\frac{\tau ^2\zeta}{2}\mathbb{E} \left[ \left\| \boldsymbol{e}^{t} \right\| _{2}^{2} \right] 
		\\
		&+\tau ^2\zeta \mathbb{E} \left[ \left< \nabla L\left( \boldsymbol{m}^t \right) ,\boldsymbol{e}^{t} \right> \right] 
		\\
		\overset{\left( b \right)}{=}&\frac{-\tau \left( 2-\tau \zeta \right)}{2}\left\| \nabla L\left( \boldsymbol{m}^t \right) \right\| _{2}^{2}+\frac{\tau ^2\zeta}{2}\mathbb{E} \left[ \left\| \boldsymbol{e}^{t} \right\| _{2}^{2} \right] 
		\\
		+&\tau \left( \tau \zeta -1 \right) \left< \nabla L\left( \boldsymbol{m}^t \right) ,\mathbb{E} \left[ \boldsymbol{e}^{t} \right] \right> 
		\\
		=&-\frac{1}{2\zeta}\left\| \nabla L\left( \boldsymbol{m}^t \right) \right\| _{2}^{2}+\frac{1}{2\zeta}\mathbb{E} \left[ \left\| \boldsymbol{e}^{t} \right\| _{2}^{2} \right] ,		
	\end{aligned}
\end{equation}
where (a) is form Assumption 2 and (b) is obtained by letting $\tau =\frac{1}{\zeta}$.
Then, the upper bound of $\mathbb{E} \left[ \left\| \boldsymbol{e}^{t} \right\| _{2}^{2} \right] $ is given as follows:

\begin{equation}\label{wg54}
	\begin{aligned}
		&\mathbb{E} \left[ \left\| \boldsymbol{e}^{t} \right\| _{2}^{2} \right] 
		\\
		=&\mathbb{E} \left[ \left\| \frac{1}{\left| \mathcal{K} \right|}\sum_{k\in \mathcal{K}}{\boldsymbol{g}_{k}^{t}}+\frac{1}{\left| \mathcal{K} \right|\nu}\boldsymbol{r}\left( t \right) -\nabla L\left( \boldsymbol{m}^t \right) \right\| _{2}^{2} \right] 
		\\
		=&\mathbb{E} \left[ \left\| \frac{1}{\left| \mathcal{K} \right|}\sum_{k\in \mathcal{K}}{\boldsymbol{g}_{k}^{t}}-\nabla L\left( \boldsymbol{m}^t \right) \right\| _{2}^{2} \right] +\mathbb{E} \left[ \left\| \frac{1}{\left| \mathcal{K} \right|\nu}\boldsymbol{r}\left( t \right) \right\| _{2}^{2} \right] 
		\\
		&+\frac{1}{\left| \mathcal{K} \right|\nu}\left< \frac{1}{\left| \mathcal{K} \right|}\sum_{k\in \mathcal{K}}{\boldsymbol{g}_{k}^{t}}-\nabla L\left( \boldsymbol{m}^t \right) ,\mathbb{E} \left[ \boldsymbol{r}\left( t \right) \right] \right> 
		\\
		\overset{\left( a \right)}{=}&\mathbb{E} \left[ \left\| \frac{1}{\left| \mathcal{K} \right|}\sum_{k\in \mathcal{K}}{\boldsymbol{g}_{k}^{t}}-\nabla L\left( \boldsymbol{m}^t \right) \right\| _{2}^{2} \right] +\frac{1}{\left| \mathcal{K} \right|^2\nu^{2}}\mathbb{E} \left[ \left\| \boldsymbol{r}\left( t \right) \right\| _{2}^{2} \right] 
		\\
		=&\mathbb{E} \left[ \left\| \frac{1}{\left| \mathcal{K} \right|}\sum_{k\in \mathcal{K}}{\boldsymbol{g}_{k}^{t}}-\nabla L\left( \boldsymbol{m}^t \right) \right\| _{2}^{2} \right] +\frac{d\sigma^2}{\left| \mathcal{K} \right|^2\nu^{2}}
		\\
		=&\mathbb{E} \left[ \left\| \frac{1}{\left| \mathcal{K} \right|}\sum_{k\in \mathcal{K}}{\boldsymbol{g}_{k}^{t}}-\frac{1}{N}\sum_{k\in \mathcal{K}}{\boldsymbol{g}_{k}^{t}}-\frac{1}{N}\sum_{k\in \mathcal{N} /\mathcal{K}}{\boldsymbol{g}_{k}^{t}} \right\| _{2}^{2} \right] +\frac{d\sigma^2}{\left| \mathcal{K} \right|^2\nu^{2}}
		\\
		=&\mathbb{E} \left[ \left\| \left( \frac{1}{\left| \mathcal{K} \right|}-\frac{1}{N} \right) \sum_{k\in \mathcal{K}}{\boldsymbol{g}_{k}^{t}}-\frac{1}{N}\sum_{k\in \mathcal{N} /\mathcal{K}}{\boldsymbol{g}_{k}^{t}} \right\| _{2}^{2} \right] +\frac{d\sigma^2}{\left| \mathcal{K} \right|^2\nu^{2}}
		\\
		\overset{\left( b \right)}{\leqslant}&\mathbb{E} \left[ \left( \left( \frac{1}{\left| \mathcal{K} \right|}-\frac{1}{N} \right) \sum_{k\in \mathcal{K}}{\left\| \boldsymbol{g}_{k}^{t} \right\| _2}+\frac{1}{N}\sum_{k\in \mathcal{N} /\mathcal{K}}{\left\| \boldsymbol{g}_{k}^{t} \right\| _2} \right) ^2 \right] 
		\\
		&+\frac{d\sigma^2}{\left| \mathcal{K} \right|^2\nu^{2}}
		\\
		%\overset{\left( c \right)}{\leqslant}&4\varpi ^2\left( \frac{1}{\left| \mathcal{K} \right|}-\frac{1}{N} \right) ^2+\frac{d\sigma^2}{\left| \mathcal{K} \right|^2\nu _{\left[ \mathcal{K} \right]}^{2}},
		\overset{\left( c \right)}{\leqslant}&4\varpi ^2\left( 1-\frac{\left| \mathcal{K} \right|}{N} \right) ^2+\frac{d\sigma^2}{\left| \mathcal{K} \right|^2\nu^{2}},
	\end{aligned}
\end{equation}
where (a) is from the fact that $\mathbb{E} \left[ \boldsymbol{r}\left( t \right) \right] =0$. Step (b) is achieved by the triangle-inequality and (c) is from Assumption 1.
By submitting (\ref{wg54}) into (\ref{wg53}) and defining $\eta =1-\frac{\varrho}{\zeta}
$, we obtain
\begin{equation}\label{wg15}
	\begin{aligned}
		&\mathbb{E} \left[ L\left( \boldsymbol{m}^{t+1} \right) -L\left( \boldsymbol{m}^* \right) \right] 
		\\
		\overset{\left( a \right)}{\leqslant}&\left( 1-\frac{\varrho}{\zeta} \right) \mathbb{E} \left[ L\left( \boldsymbol{m}^t \right) -L\left( \boldsymbol{m}^* \right) \right] 
		\\
		&+\frac{1}{2\zeta}\left( 4\varpi ^2\left( 1-\frac{\left| \mathcal{K} \right|}{N} \right) ^2+\frac{d\sigma ^2}{\left| \mathcal{K} \right|^2\nu^{2}} \right) 
		\\
		\leqslant& \left( 1-\frac{\varrho}{\zeta} \right) ^t\mathbb{E} \left[ L\left( \boldsymbol{m}^0 \right) -L\left( \boldsymbol{m}^* \right) \right] 
		\\
		&+\frac{1}{2\zeta}\left( 4\varpi ^2\left( 1-\frac{\left| \mathcal{K} \right|}{N} \right) ^2+\frac{d\sigma ^2}{\left| \mathcal{K} \right|^2\nu^{2}} \right) \sum_{i=0}^{t-1}{\left( 1-\frac{\varrho}{\zeta} \right) ^i}
		\\
		\leqslant& \eta ^t\mathbb{E} \left[ L\left( \boldsymbol{m}^0 \right) -L\left( \boldsymbol{m}^* \right) \right] 
		\\
		&+\frac{1}{2\zeta}\left( 4\varpi ^2\left( 1-\frac{\left| \mathcal{K} \right|}{N} \right) ^2+\frac{d\sigma ^2}{\left| \mathcal{K} \right|^2\nu^{2}} \right) \frac{1-\eta^t}{1-\eta},
	\end{aligned}
\end{equation}	
where (a) is form Assumption \ref{Polyak}. 
Finally, by replacing $t+1$ with $T$, we complete the proof.

\bibliographystyle{IEEEtran}
\bibliography{FLbib.bib}
\end{document}